\documentclass[aps,prl,superscriptaddress,twocolumn,floatfix,preprintnumbers]{revtex4}
 
\usepackage{amsmath}
\usepackage{graphicx,tabularx,epsfig}
\usepackage{comment}

\def\beq{\begin{equation}}
\def\eeq{\end{equation}}
\def\be{\begin{equation}}
\def\ee{\end{equation}} 
\def\bea{\arraycolsep .1em \begin{eqnarray}}
\def\eea{\end{eqnarray}}

\def\s0#1#2{\mbox{\small{$ \frac{#1}{#2} $}}}
\def\0#1#2{\frac{#1}{#2}}

\begin{document}
\graphicspath{{FIGURE/}} 
 
\title{Stability, Higgs Boson Mass and New Physics}
 
\author{Vincenzo Branchina and Emanuele Messina}
\affiliation{Department of Physics, University of Catania 
and INFN, Sezione di Catania, Via Santa Sofia 64, I-95123 
Catania, Italy }  

\begin{abstract}
Assuming that the particle with mass $\sim 126$ GeV discovered 
at LHC is the Standard Model Higgs boson, we find that the 
stability of the EW vacuum strongly depends on new physics 
interaction at the Planck scale $M_P$, despite of the fact 
that they are higher-dimensional interactions, apparently 
suppressed by inverse powers of $M_P$. In particular, for 
the present experimental values of the top and Higgs masses,  
if $\tau$ is the lifetime of the EW vacuum, new physics can 
turn $\tau$ from $\tau >> T_U$ to $\tau << T_U$, where $T_U$ 
is the age of the Universe, thus weakening the  
conclusions of the so called meta-stability scenario.  
\end{abstract}

\maketitle
 
{\it Introduction.---} 
When the particle with mass $\sim 126$ GeV discovered at 
LHC\,\cite{atlas, cms} is identified 
with the Standard Model (SM) Higgs boson, serious and 
challenging questions arise. Among them, the vacuum stability 
issue.
The Higgs effective potential $V_{eff}(\phi)$ bends down 
for values of $\phi$ larger the EW minimum, an instability 
due to top loop-corrections. By requiring stability, lower 
bounds on the Higgs 
mass $M_H$ were found\,\cite{cab, sher, jones, sher2, alta, 
quiro,shapo}. 

\begin{figure}[t]
\vskip-4mm
\includegraphics[width=.44\textwidth]{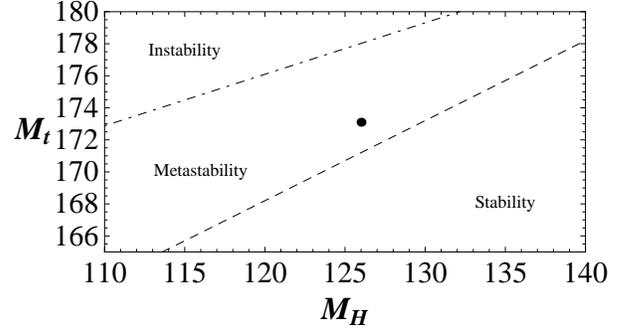}\vskip-4mm
\caption{In this picture we repeat the analysis 
of\,\cite{isido, isiuno, isidue}, which is done in the 
absence of new interactions at the Planck scale. The 
$M_H-M_t$ plane is divided in three sectors: absolute 
stability, metastability and instability regions. 
The dot indicates $M_H\sim 126$ GeV 
and $M_t\sim 173.1$ GeV.
}
\label{bounn}\vskip-.6cm
\end{figure}

A variation on this picture is the so called metastability 
scenario\,\cite {sher, isido, isiuno, isidue}. 
For $\phi$ much larger than $v$ (location of the EW minimum), 
$V_{eff}(\phi)$ develops a new minimum at $\phi_{min}^{(2)}$. 
When $M_H$ and $M_t$ are such that $V_{eff}(v) < 
V_{eff}(\phi_{min}^{(2)})$, the EW minimum is {\it stable}, 
otherwise it is a false vacuum that should decay into the true 
vacuum (at $\phi_{min}^{(2)}$) in a 
finite amount of time. Depending on the values of $M_H$ 
and $M_t$, the lifetime $\tau$ of the EW vacuum can be larger 
or smaller than the age 
of the Universe $T_U$. For $\tau > T_U $, we may well live in 
the metastable EW minimum. This is the metastability 
scenario. 

The aim of this Letter is to study the influence 
of new physics interactions (at the Planck scale) on 
$\tau$. Tree level and quantum fluctuation contributions
are taken into account. In this Letter, however, we limit 
ourselves to consider the quantum corrections  
from the Higgs sector only. This is sufficient to  
illustrate our main point. The complete analysis is 
left for a forthcoming paper. 

Let us begin with fig.\,\ref{bounn}, where we repeat the  
usual analysis\,\cite{isido, isiuno, isidue} and draw the 
phase diagram in the $M_H-M_t$ plane. The latter is divided into 
three different sectors: an {\it absolute stability} region 
($V_{eff}(v) < V_{eff}(\phi_{min}^{(2)})$), a {\it  metastability} 
region ($\tau > T_U $) and an {\it instability} region  
($\tau < T_U $). The dashed line separates the stability and 
the metastability sectors and is obtained for 
$M_H$ and $M_t$ such that $V_{eff}(v) = 
V_{eff}(\phi_{min}^{(2)})$. The dashed-dotted line 
separates the metastability and the instability regions and is 
obtained for $M_H$ and $M_t$ such that $\tau = T_U $.
For $M_t \sim 173.1$ GeV and $M_H \sim 
126$ GeV, the SM lies within the metastability region. 
It is then concluded that the present experimental values of 
$M_H$ and $M_t$ allow for a Standard 
Model valid all the way up to the Planck scale.  

Let $V_{eff}(\phi)$ be normalized so to vanish at 
$\phi=v$. At a much larger value $\phi=\phi_{inst}$, 
$V_{eff}(\phi_{inst})$ vanishes again (for 
$M_H \sim 126$ GeV, $M_t \sim 173.1$ GeV, this happens for
$\phi_{inst} \sim 10^{10}$ GeV). For $\phi > \phi_{inst}$,
the potential becomes negative, later developing a new minimum. 

It is assumed that the actual behavior of $V_{eff}(\phi)$ 
for $\phi$ beyond $\phi_{inst}$ has no impact on $\tau$. 
More precisely, it is stated that even if 
$V_{eff}(\phi)$ at $\phi=M_P$ is still negative 
(and the new minimum forms at a scale much larger than $M_P$),
new physics interactions 
around the Planck scale must stabilize the potential 
(eventually bringing the new minimum around $M_P$), but  
$\tau$ does not depend on the detailed 
form of $V_{eff}(\phi)$ beyond $\phi_{inst}$\,\cite{isido}. 

In this respect, it is worth to note that for 
$M_H \sim 126$ GeV and  $M_t\sim 173.1$ GeV, not only 
the effective potential at the Planck scale is negative but also   
it continues to go down beyond $M_P$. The new minimum is formed at 
$\phi_{min}^{(2)} \sim 10^{31}$ GeV (see fig.\ref{instab}). 

Note also that the instability of the effective potential 
occurs for very large values of $\phi$,  
($\phi_{inst} \sim 10^{10}$ GeV). In 
this range, $V_{eff}(\phi)$ is well approximated by keeping 
only the quartic term\,\cite{sher2}. 
Accordingly, following\,\cite{cole1, cole2}, the tunneling 
time $\tau$ is 
computed by considering the bounce solutions to the euclidean 
equation of motion for the potential $V(\phi)= 
\frac{\lambda}{4}\phi^4$ with negative $\lambda$, 
a good approximation in this range.

\begin{figure}[t]
\vskip-4mm
\includegraphics[width=.44\textwidth]{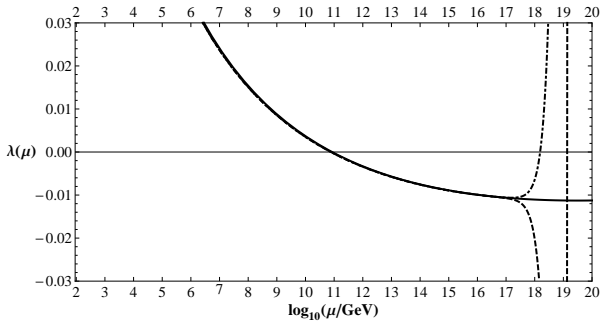}\vskip-4mm
\caption{For $M_H \sim 126$ GeV and $M_t\sim 173.1$ GeV,
the running of $\lambda(\mu)$ as determined by $SM$ model 
interactions only (solid line) and in the presence of 
$\lambda_6$ and $\lambda_8$. Dashed-dotted line:  
$\lambda_6(M_P)=1$ and $\lambda_8(M_P)=0.5$. Dashed line:
$\lambda_6(M_P)=-2$ and $\lambda_8(M_P)=2.1$. Clearly, 
the tree lines 
coincide for values of $\mu$ below the Planck scale.
}
\label{lam}\vskip-.6cm
\end{figure} 

{\it Lifetime of the EW vacuum.---} 
In order to study the impact of new physics 
interactions at the Planck scale, we add
two higher dimension operators $\phi^6$ and 
$\phi^8$ to the SM Higgs 
potential:
\begin{eqnarray}\label{newpot}
V(\phi)&=&\frac{\lambda}{4}\phi^4
+\frac{\lambda_6}{6}\frac{\phi^6}{M_P^2}
+\frac{\lambda_8}{8}\frac{\phi^8}{M_P^4}\,\,.
\end{eqnarray} 
Naturally, we can also consider higher dimensional operators. 
However, the examples we are going to study (with different 
choices of $\lambda_6$ and $\lambda_8$) are sufficient 
to illustrate some interesting cases we can face when new 
physics interactions at the Planck scale are considered.

The influence of $\lambda_6$ and $\lambda_8$ on the 
RG flow of the quartic coupling $\lambda(\mu)$, for values 
of $\mu$ below $M_P$, is 
negligible (see fig.\,\ref{lam}). 
The RG functions for the SM parameters at the two-loop 
level (with the corresponding boundary conditions) can be 
found, for instance, 
in\,\cite{isiuno,riotto}. Further (slight) 
improvement is obtained by considering three-loops 
contributions\,\cite{isiuno,isidue,bla}.

Let us consider now
two different representative cases. For $\lambda_6 (M_P)=-2$ 
and $\lambda_8 (M_P)=2.1$, the potential is given by the dashed 
line of fig.\,\ref{instab}. Due to the large range of scales 
involved, the plot is done in a double logarithmic scale. As 
$\lambda_6$ is negative, when $\phi$ approaches 
$M_P$, $V_{eff}^{new}(\phi)$,which is the renormalization group 
improved effective potential in the presence of $\lambda_6$ 
and $\lambda_8$, bends down 
much steeply than $V_{eff}(\phi)$ and forms a new minimum 
at about $\phi_{min}^{(2)}\sim M_P$. This is clear from 
the zoom around the Planck scale
in panel (b) of fig.\,\ref{instab}. 

The second case we consider is when $\lambda_6$ and 
$\lambda_8$ are both positive. For $\lambda_6 (M_P)=1$ 
and $\lambda_8 (M_P)=0.5$, the potential is given by the 
dotted-dashed line of fig.\,\ref{instab}. 
As $\lambda_6$ is positive, 
when $\phi$ approaches $M_P$ the potential 
$V_{eff}^{new}(\phi)$ lies above (rather than below) 
$V_{eff}(\phi)$. 

In both cases, the potential is stabilized at the Planck 
scale by new physics terms. 
However, 
it is commonly believed that, although 
such a stabilization has to take place, the presence 
of new physics interactions has no impact on the EW 
vacuum lifetime\,\cite{isido}.  
We shall see that this is not generically true.
When $V_{eff}^{new}(\phi)$ lies above 
$V_{eff}(\phi)$, which in our example is realized with 
$\lambda_6(M_P) > 0$ and $\lambda_8(M_P) > 0$, $\tau$
is almost insensitive to the presence of these new terms. 
On the contrary, when $V_{eff}^{new}(\phi)$ lies below 
$V_{eff}(\phi)$, which in our example is realized with 
$\lambda_6(M_P) < 0$ and $\lambda_8(M_P) > 0$,  $\tau$ 
strongly depends on new physics. 

The tunneling time $\tau$ is given 
by\,\cite{cole1,cole2},\,\cite{isido}:
\begin{eqnarray}\label{tunneling}
\frac{1}{\tau} = T_U^{3}\,\frac{S[\phi_b]^2}{4\pi^2}
\left|\frac{ {\rm det'}\left[-\partial^2+V''(\phi_b)\right]}
{\mbox{det}\left[-\partial^2+V''(v)\right]}\right|^{-1/2} e^{-S[\phi_b]} 
\end{eqnarray}
where $\phi_b(r)$ is the $O(4)$ bounce solution to the euclidean 
equation of motion ($r^2=x_\mu x_\mu $), $S[\phi_b]$ the action for 
the bounce,
$\left[-\partial^2+V''(\phi_b)\right]$ the fluctuation operator
around the bounce ($V''$ is the second derivative of $V$
with respect to $\phi$). The prime in the ${\rm det^{'}}$ 
means that in the computation of the determinant the zero modes 
are excluded and $\frac{S[\phi_b]^2}{4\pi^2}$ comes from the 
translational zero modes.

\begin{figure*}[t]
\begin{minipage}[c]{0.49\textwidth}
\includegraphics[width=0.75\textwidth,angle=0]{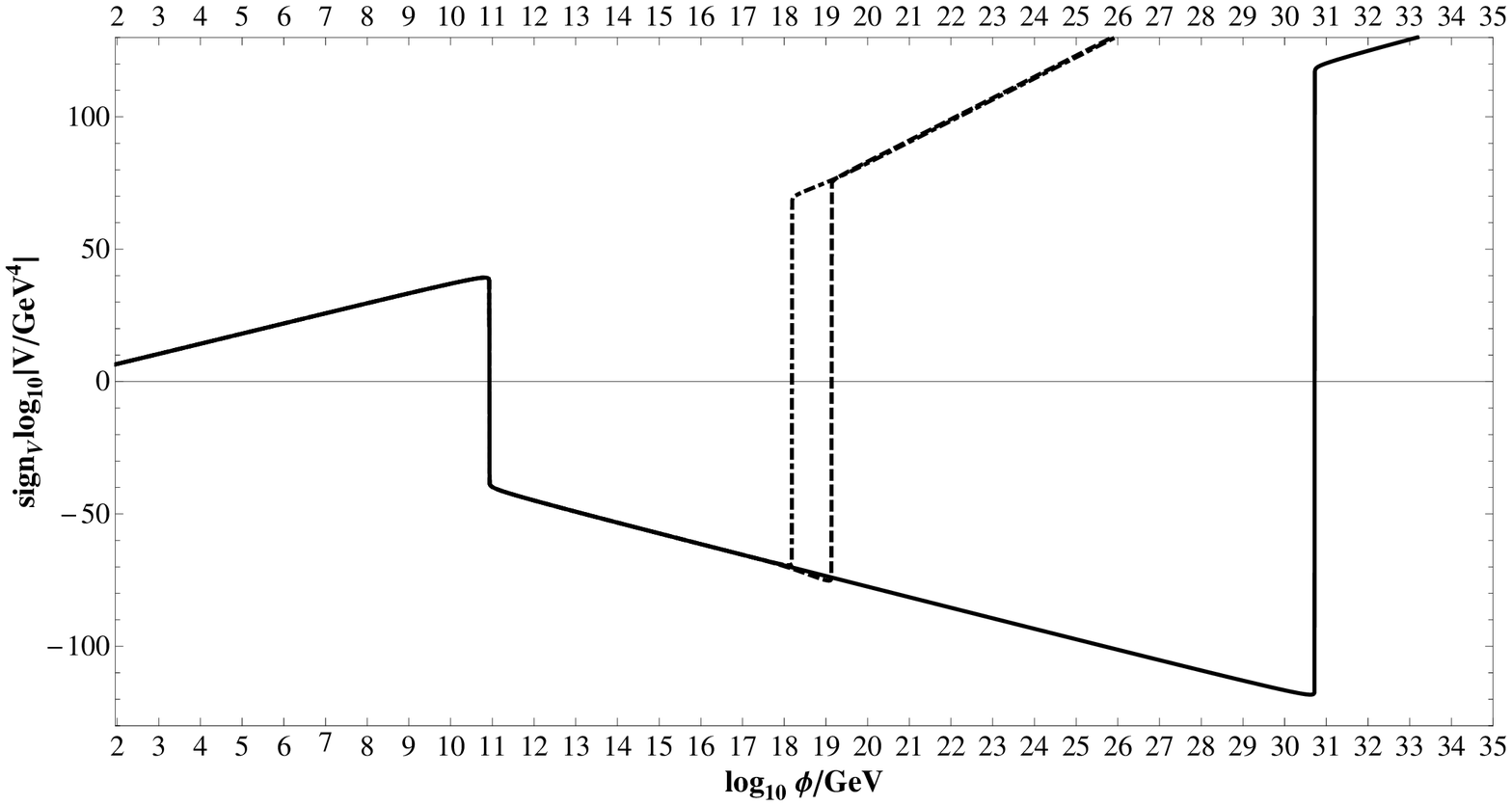}
\\\centering (a)
\end{minipage}
\begin{minipage}[c]{0.49\textwidth}
\includegraphics[width=0.75\textwidth,angle=0]{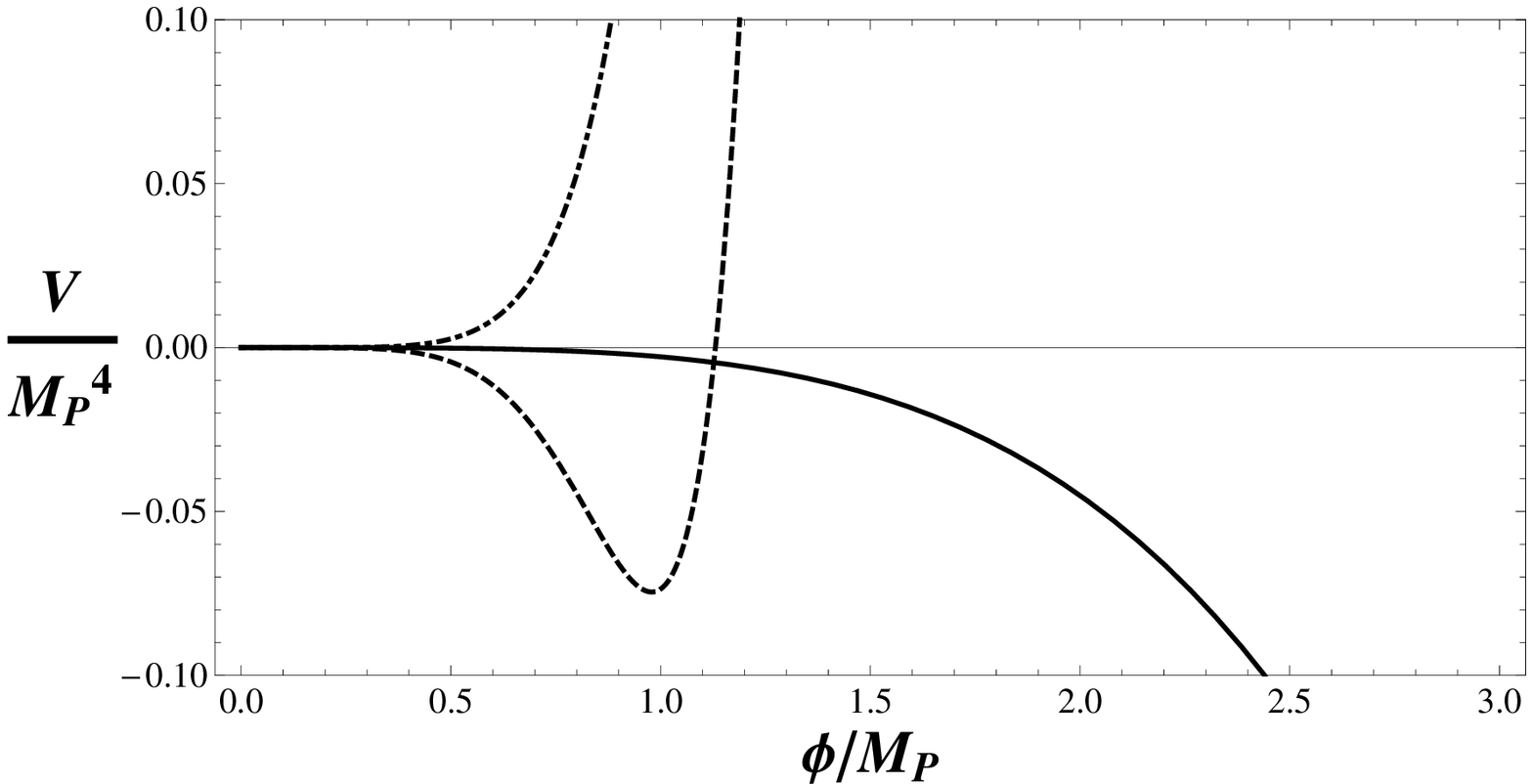}
\\\centering (b)
\end{minipage}
\caption{(a) The effective potential 
$V_{eff}(\phi)$ (solid line) for $M_H \sim 126$ GeV and 
$M_t\sim 173.1$ GeV. Note that the new minimum forms at 
$\phi_{min}^{(2)} \sim 10^{31}$ GeV. For the same values of $M_H$ and 
$M_t$, $V_{eff}^{new}(\phi)$ for $\lambda_6=-2$ and 
$\lambda_8=2.1$ (dashed line); $V_{eff}^{new}(\phi)$ for 
$\lambda_6=1$ and $\lambda_8=0.5$ (dashed-dotted line). In order
to include such a vaste range of scales, a log-log plot has been 
considered.  
(b) Zoom of the panel (a) figure near the Planck scale.  
$V_{eff}(\phi)$, $V_{eff}^{new}(\phi)$ and $\phi$ are normalized
to Planck units (in this range no log-log plot is needed).  
$V_{eff}^{new}(\phi)$ for $\lambda_6 <0$ bends down steeply and 
forms a new minimum at $\phi_{min}^{(2)}=0.979 M_P$. 
With $\lambda_6 > 0$, $V_{eff}^{new}(\phi)$ falls down less
steeply than $V_{eff}(\phi)$ and in the picture we cannot
resolve the minimum which forms at $\phi=0.119 M_P$.} \label{instab}
\end{figure*}

Let us compute $\tau$ 
for the potential of Eq.\,(\ref{newpot}) with 
$\lambda_6(M_P) =-2$ and $\lambda_8(M_P) =2.1$. 

When $V_{eff}(\phi)$ (the usual SM Higgs potential without 
new interaction terms) is computed within the $\overline {MS}$ 
scheme and the renormalization scale $\mu$ is chosen 
to coincide with the inverse of the bounce size $R_{max}$ that 
maximizes the tunneling probability\,\cite{isido}, we have that 
$\mu=2\,R_{max}^{-1}\,e^{-\gamma_E} =1.32\,\cdot \, 10^{17}$ GeV
($\gamma_E$ is the Euler 
gamma) and the coupling constant 
$\lambda(\mu)$ is $\lambda(\mu)\simeq-0.014$.  

For our potential, we find that, up 
to the scale $\eta\simeq 0.780 M_P$, it is very well approximated 
by an upside down quartic parabola, 
$V_{eff}^{new}(\phi)=\frac{\lambda_{eff}}{4}\phi^4$, with  
$\lambda_{eff}=\lambda+\frac{2}{3}\lambda_6\frac{\eta^2}{M_P^2}
+\frac{1}{2}\lambda_8\frac{\eta^4}{M_P^4} \simeq -0.437$. 
For $\phi > \eta$, 
$V_{eff}^{new}(\phi)$ bends down very steeply (see fig. 3 (b)), 
eventually 
creating a new minimum very close to $M_P$ : 
$\phi_{min}^{(2)}=0.979 M_P$. Therefore, for values of $\phi$
larger than (but close to) $\eta$,
$\phi \gtrsim \eta$, it can be linearized and we get:

\noindent
$V(\phi)=\left[\frac{\lambda_{eff}}{4}
\eta^4-\frac{\lambda_{eff} \eta^3}{\gamma}
\left(\phi-\eta\right)\right]$,  where  
\begin{eqnarray}\label{gamma}
\gamma= -\, \lambda_{eff} \,\,\eta^3 \, \left(\lambda\eta^3
+\lambda_6\frac{\eta^5}{M_P^2}+\lambda_8\frac{\eta^7}
{M_P^4}\right)^{-1}\,\, .
\end{eqnarray}

Interestingly, in order to compute $\tau$, this is all what we 
need to know\,\cite{wein}. Moreover, the parameter 
$\gamma$ plays an essential role in determining when new 
physics interactions influence $\tau$. 

The euclidean equation of motion admits the following 
bounce solution:
\begin{eqnarray}\label{boun}
\phi_b(r)=\left\{\begin{array}{cc}
2 \eta -\eta^2 \sqrt{\frac{|\lambda_{eff}|}{8}}
\frac{r^2+\overline R^2}{\overline R} & \qquad
0<r<\overline r\\
\sqrt{\frac{8}{|\lambda_{eff}|}}\frac{\overline R}{r^2+\overline R^2}
& \qquad
r>\overline r
\end{array}\right.
\end{eqnarray}
where 
\begin{equation}\label{erre}
\overline r^2=\frac{8\gamma}{\lambda_{eff} \eta^2}(1+\gamma)
\,\,\,\,\,\,\,\, , \,\,\,\,\,\,\,\, 
\overline R^2=\frac{8}{|\lambda_{eff}|}\frac{\gamma^2}{\eta^2}\,.
\end{equation}
From Eq.\,$(\ref{erre})_1$ we see that the solution 
(\ref{boun}) exists only for  
$-1 < \gamma < 0$.  \, $\overline R$ is 
the size of the bounce (\ref{boun})
and the action at $\phi_b$ is: 
\begin{eqnarray}\label{bounceone}
S[\phi_b]= (1 - (\gamma + 1)^4)\,\,\frac{8\pi^2}{3|\lambda_{eff}|}\,.
\end{eqnarray}

There are also other bounce solutions: 
\begin{eqnarray}\label{bouncetwo}
\phi_b^{(2)}(r)=\sqrt{\frac{2}{|\lambda_{eff}|}}\frac{2\,R}{r^2+R^2}
\end{eqnarray} 
where $R$, the size of these bounces, can take any value
in the range  
$\sqrt{\frac{8}{|\lambda_{eff}|}}\frac{1}{\eta} < R < \infty$.
A numerical analysis (presented in detail in a forthcoming paper)
shows that these latter solutions are related to the 
approximation that we are considering for $V(\phi)$. 
Nevertheless, for $|\phi| << M_P$ (that in turn means for 
very large values of $R$), configurations of the kind given in 
Eq.\,(\ref{bouncetwo}), with $\lambda_{\rm eff}$ replaced with 
$\lambda$, are good approximate solutions of the (Euclidean) 
equation of motion, and, in principle, should be taken into 
account in the computation of $\tau$. 
Their action is degenerate with $R$ and is 
\begin{eqnarray}\label{newactione}
S=\frac{8\pi^2}{3|\lambda|}\,.
\end{eqnarray} 

However, even if for a moment we limit ourselves to the tree 
level contribution only, from Eqs.\,(\ref{tunneling}),   
(\ref{bounceone}), and (\ref{newactione}) we see that  
for those values of $\gamma$ such 
that the solution (\ref{boun}) exists ($-1<\gamma<0$), the 
contribution to the tunneling probability coming from the 
bounces (\ref{bouncetwo}) (with $\lambda_{\rm eff}$ replaced with 
$\lambda$)
is exponentially suppressed with respect to the contribution of 
(\ref{boun}). For $\lambda$, 
$\lambda_6(M_P)$, $\lambda_8(M_P)$ and 
$\eta$ given above we have: $\gamma\simeq -0.963$. 

Let us compute now the fluctuation determinant 
in Eq.(\ref{tunneling}) for the bounce (\ref{boun})
and for $\lambda_6(M_P)=-2$ and $\lambda_8(M_P)=2.1$,
which is the case of interest for us.
 
Due to radial symmetry, $V''(\phi_b)$ in 
$\left[-\partial^2+V''(\phi_b)\right]$ only depends on $r$. 
Following \cite{yaglom}, the 
logarithm of the fluctuation determinant is  
obtained (see below for some specifications) as:  
\be \label{yaglom}
\log\left(\frac{\rm det' (-\partial^2+V''(\phi_b))}
{\rm det (-\partial^2)}\right)^{1/2}=
\frac{1}{2}\sum_{l= 0}^{\infty} (l+1)^2\ln \rho_l\\
\ee 
\be
{\rm where}\,\,\,\,\,\,\rho_l =\lim_{r\rightarrow \infty} 
\rho_l(r)\,\,\,\,\,\,\,\,\,\,\,\,\,\,\,\,\,\,\,\,\,\,\,\,
\ee
and each $\rho_l(r)$ is solution of the differential equation:
\begin{eqnarray}\label{yagloeq}
\rho_l''(r)+\frac{\left(2l+d-1\right)}
{r}\rho_l'(r)-V''(\phi_b(r))\rho_l(r)=0
\end{eqnarray}  
with boundary conditions $\rho_l(0)=1$ and $\rho_l'(0)=0$.
($\rho_l''(r)$ is the second derivative of $\rho_l(r)$ w.r.to $r$,...).

Eq.\,(\ref{yaglom}) is ill defined in three respects. 
The eigenvalue related to $l=0$ is a negative mode ($\rho_0<0$), 
while the $l=1$ modes correspond to the four translational 
zero modes. We exclude the $l=0$ and $l=1$ modes 
from the above sum. They can be separately treated in 
a standard way\,\cite{yaglom,dunne}. Finally, the sum in 
Eq.\,(\ref{yaglom}) is divergent. This is the usual UV 
divergence and we know how to take care of it through 
renormalization\,\cite{dunne}.

Let us consider now Eq.\,(\ref{yagloeq}) for $l>1$. 
We can easily solve this equation numerically for each value 
of $l$ (for increasing $l$, the $\rho_l$'s rapidly converge to
one). Following\,\cite{dunne}, the $\overline {MS}$ renormalized sum 
in Eq.\,(\ref{yaglom}) is given by: 
\begin{eqnarray}\label{renorm}
&&\left[\frac{1}{2}\sum_{l>1}^{\infty} 
(l+1)^2\ln \rho_l\right]_{r}= 
\frac{1}{2}\sum_{l>1}^{\infty}(l+1)^2\ln\rho_l\nonumber\\
-&\frac{1}{2}&
\sum_{l=0}^{\infty}(l+1)^2\left[\frac{\int^{\infty}_0 dr r
V''}{2(l+1)}-\frac{\int^{\infty}_0 dr r^3
(V'')^2}{8(l+1)^3}\right]\nonumber\\
-&\frac{1}{8}&\int^{\infty}_0 dr r^3 (V'')^2
\left[\ln\left(\frac{\mu r}{2}\right)+\gamma_E+1\right]\,,
\end{eqnarray}
where $\mu$ is the renormalization scale. We then get:
\begin{eqnarray}\label{ya}
\left[\frac{1}{2}\sum_{l>1}^{\infty} 
(l+1)^2\ln \rho_l\right]_{r}= -2.49 - 5.27\,\, 
{\rm ln } \left(\frac{1.48\mu}{M_P}\right)\, . \,\, 
\end{eqnarray} 

This result is obtained by truncating the sum to a 
value of $l$ where it shows saturation (standard 
renormalization procedure). Strictly speaking, the ``angular 
momentum'' cutoff $L$ in this sum is given by 
$L={\overline R} M_P$, which, from Eq.\,$(\ref{erre})_2$ is 
$L \sim 5$.
However, the series in Eq.\,(\ref{renorm}) converges 
very fast. Even truncating it to $l=5$ we get a less than 
3 percent difference with the result of Eq.\,(\ref{ya}). 
The standard renormalization procedure is 
then well justified.

For $l=0$, $\rho_0$ has to be replaced with its absolute 
value\,\cite{dunne}. Solving Eq.\,(\ref{yagloeq}), 
we find that its contribution to the sum in 
Eq.\,(\ref{yaglom}) is: $\frac12\ln |\rho_0| = -0.806$.

Finally, the contribution of the zero modes ($l=1$) is 
also obtained in a standard manner\,\cite{dunne}. The solution
of Eq.\,(\ref{yagloeq})  for $l=1$ vanishes in the 
$r \to \infty$ limit: $\rho_1=0$.
Actually,  $\rho_1$ has to be replaced with 
$\rho_1'$, defined as:
\be
\rho_1'=\lim_{k \to 0}\frac{\rho_1^{k}}{k^2}
\ee 
where $\rho_1^{k}$ is obtained by solving Eq.\,(\ref{yagloeq}) 
with $V''(\phi_b)$ replaced by $V''(\phi_b) + k^2$.
Note that $\rho'_1$ has the dimension of a length square and 
is given in terms of $\overline R$, the size of the bounce 
(\ref{boun}). The zero modes contribution to the sum in 
Eq.\,(\ref{yaglom}) finally is: 
$\frac12\,\, 4\,\,\ln \rho'_1 = 2 \,{ln}\,(0.0896 \overline R^2)$. 

For the purposes of comparing our results 
(from $V_{eff}^{new}(\phi)$) with those obtained 
with $V_{eff}(\phi)$, it is useful to choose the same 
renormalization scale as before,  
$\mu=1.32\,\cdot\, 10^{17}$ GeV. Then, collecting the 
different results, from Eq.\,(\ref{tunneling}) we find: 
\be\label{newtau}
\tau= 5.45 \cdot 10^{-212}\,\, T_U,
\ee
a ridiculously small fraction of a second! 

This result is at odds with what is shown in 
fig.\,(\ref{bounn}), where for $M_H\sim 126$ GeV and 
$M_t\sim 173.1$ GeV, the EW vacuum lies inside 
the metastability region, close to the stability line and shows 
that the phase diagram of fig.\,\ref{bounn} has to be reconsidered. 
Actually, when the EW vacuum lifetime for these values of $M_H$ and 
$M_t$ is computed in the absence of new physics interactions,
we have:
\be
\tau= 1.49 \cdot 10^{714}\,\, T_U\,. \label{tauloro}
\ee  
Accordingly, the EW vacuum would 
be an extremely long-lived metastable state. This is why it is 
often stated that, for the present experimental values of
$M_H$ and $M_t$, the SM is an effective theory that is valid 
all the way up to the Planck scale. 

Eq.\,(\ref{newtau}) shows that this is not 
generically true. As a result of the presence of new physics 
interactions, the EW vacuum may turn from a very long-lived 
metastable state to a highly unstable one. As we have 
already seen, in fact, when 
$V_{eff}^{new}(\phi)$ lies above $V_{eff}(\phi)$,\, 
 $\tau$ is not dramatically affected by new physics.
On the contrary, when $V_{eff}^{new}(\phi)$ lies below 
$V_{eff}(\phi)$, the $UV$ completion of the Standard Model 
has a very strong impact on $\tau$, turning it from 
$\tau>>T_U$ to $\tau<<T_U$.

{\it Conclusions and outlook.---}
In this Letter we show that the lifetime $\tau$ of the EW 
vacuum strongly depends on new physics. The metastability 
scenario (which is based on the assumption that $\tau$ does 
not depend on new physics) and the whole phase structure 
of fig.\,(\ref{bounn}) have to be entirely reconsidered. 

Clearly, when the quantum fluctuations from  
other sectors of the $SM$ are taken into account, 
the specific value of $\tau$ in Eq.\,(\ref{newtau}) is modified.
However, this does not change the core result of the present analysis,
namely the huge influence of new physics on $\tau$. 

A very important outcome of our result is 
that it poses constraints on possible candidates to 
the $UV$ completion of the $SM$. In this respect, we note 
also that a similar
analysis can be done when the new physics scale 
lies below (even much below) the Planck scale.

Finally, we note that the considerations developed in this 
Letter should be relevant for related scenarios, Higgs potential 
with two degenerate minima\,\cite{niel} and Higgs driven 
inflation scenarios\,\cite{ber,infla}. In all of these cases, 
in fact, the physical scale relevant to the involved mechanism 
is dangerously close to the Planck scale and we expect high 
sensitivity to new physics interactions.

\end{document}